# Using Synchronization for Prediction of High-Dimensional Chaotic Dynamics

Adam B. Cohen[1,2], Bhargava Ravoori[1,2], Thomas E. Murphy[1,3] and Rajarshi Roy[1,2,4]

[1]Institute for Research in Electronic and Applied Physics, University of Maryland, College Park, Maryland 20742, USA
[2]Department of Physics, University of Maryland, College Park, Maryland, 20742 USA
[3]Department of Electrical and Computer Engineering, University of Maryland, College Park, Maryland 20742, USA
[4]Institute for Physical Sciences and Technology, University of Maryland, College Park, Maryland 20742, USA

We experimentally observe the nonlinear dynamics of an optoelectronic time-delayed feedback loop designed for chaotic communication using commercial fiber optic links, and we simulate the system using delay differential equations. We show that synchronization of a numerical model to experimental measurements provides a new way to assimilate data and forecast the future of this time-delayed high-dimensional system. For this system, which has a feedback time delay of 22 ns, we show that one can predict the time series for up to several delay periods, when the dynamics is about 15 dimensional.

The question of how to predict the future of a dynamical system with time delay is of interest in many applications [1–2]. In chaotic encrypted communication systems, the predictability is related to how difficult it would be for an eavesdropper to intercept a message. A better understanding of predictability in these systems could guide the development of new strategies to improve security, such as periodically changing the system parameters or protocols to avoid interception [3]. Prediction and anticipation are thought to underlie the process of image recognition and motion tracking in the retina [4]. In networked sensor arrays designed to detect spatiotemporal disturbances, prediction methods could enable efficient acquisition and incorporation of data from multiple sensors [5]. In biomedical treatment, prediction models could lead to improved strategies for adjusting drug dosage and delivery or physiological control [6].

Recent studies on prediction address system identification and model development [7], as well as the use of anticipated synchronization between coupled identical systems [8]. In this work, we demonstrate that synchronization of a numerical model to an experimentally measured waveform allows us to both forecast the future dynamics of a high-dimensional system and estimate the local maximum Lyapunov exponent and its distribution. The inverse of the maximum Lyapunov exponent defines the prediction horizon – the time over which the system behavior can be forecast.

The optoelectronic system studied is shown in Figure 1. A similar system was used by Argyris *et al*. as a transmitter and receiver for a high-speed chaotic communication demonstration over a commercial fiber optic channel [9]. If the high- and low-pass filters are approximated with a single-pole response, the system can be modeled by the delay differential equations [10]

$$
\begin{aligned}
\tau_L \frac{dx(t)}{dt} &= -\left(1 + \frac{\tau_L}{\tau_H}\right) x(t) - y(t) - \beta \cos^2[x(t-\tau) + \phi_o] \\
\tau_H \frac{dy(t)}{dt} &= x(t)
\end{aligned}
\tag{1}
$$

where $x(t)$ is a normalized output signal representing the voltage applied to the modulator, $\beta$ is a dimensionless constant that describes the feedback strength of the loop, $\tau$ is the net time delay of the feedback signal, and $\phi_o$ represents the bias point of the Mach-Zehnder modulator. $\tau_L$ and $\tau_H$ are time constants describing the low-pass and high-pass filters, *i.e.* the low-pass cut-off frequency is $2\pi/\tau_L$ and the high-pass cut-on frequency is $2\pi/\tau_H$.

This simple model however fails to provide an accurate description of the dynamics of the experimental system considered in this study. The supplementary material [11] explains how this simple model can be generalized to incorporate arbitrary band-pass filters, such as the 7th order Butterworth filters that where actually employed in our system. The prediction techniques described here are general and could also be applied to photonic monolithic integrated devices recently developed for chaotic communications [12].

Figure 2(a) shows experimental time traces from this nonlinear optoelectronic loop, for five different values of the feedback strength, together with simulated data obtained by numerically integrating the equations of motion presented in the supplementary material [11]. In the experimental measurements, the feedback strength was varied by adjusting the laser power input to the modulator.

Figure 2(b) shows the bifurcation diagram constructed from experimental measurements in comparison to the one obtained from numerical simulations. Figure 2(c) plots the Lyapunov dimension of the system as a function of the feedback strength $\beta$, calculated using the Kaplan-Yorke conjecture [13] after computing the Lyapunov spectrum from a linearized numerical model [14]. One feature of time-delay systems is the variability of the dimensionality and its dependence on parameter values. The supplementary material [11] provides details of a linearized model used to calculate the Lyapunov spectrum and dimension.

One of the reasons prediction proves challenging is that it is difficult to directly measure and incorporate all of the relevant variables of the model [2]. For example, in our experimental system, we measure the light intensity with a digital oscilloscope only at certain sampling rates and with limited precision. The question of how to establish all of the initial values needed for the numerical modeling and prediction poses a serious problem.

Our solution utilizes the phenomenon of synchronization between nonlinear dynamical systems [15]. As depicted in Figure 3(a), the measured experimental data, denoted $x_I(t)$, is fed into the numerical model in place of the feedback signal**.** This process of data assimilation allows the multiple variables and dimensional degrees of freedom in the model to adapt in response to the input time series. The portion of the time series before $t = 0$ in Figure 3(a) shows a close correspondence between the experimental and numerical traces, indicating that open-loop synchronization has been achieved. At $t = 0$, the input of values from the experimental measurements is terminated, and the model is numerically integrated to forecast the future dynamics of the system. The experimental time series actually measured for this period is retained for comparison. Figure 3(b) displays the absolute separation between the prediction and experiment, showing an initial divergence that eventually saturates because of the finite amplitude of the attractor. By fitting an exponential relation to this initial divergence, one determines the local maximum Lyapunov exponent and its inverse, the prediction horizon.

The degree to which the signals synchronize depends on many factors, including the accuracy of the numerical model and the noise level of the experimental data. In order to ensure the best possible predictions, the value of the parameter $\beta$ in the numerical model was adjusted to minimize the open loop synchronization error (time-averaged RMS difference). The synchronization error remains within ±5% of the minimum value for a deviation of about ±7% from the optimal value of $\beta$.

The same analysis can be carried out using independently simulated data in place of experimental data. Because the simulated data were obtained using equations identical to those used in the prediction model, the two signals achieve a much closer initial synchrony. The dashed curve in Figure 3(c) shows the absolute difference between two data sequences that were both obtained from simulation, and their subsequent exponential divergence after $t = 0$.

The average Lyapunov exponent of the system is calculated by applying this method at many different points on the attractor, *i.e.*, by closing the switch at different points in time and fitting the subsequent exponential divergence. Figure 4(a) shows the average Lyapunov exponent obtained as a function of the feedback strength $\beta$. The open circles show the average Lyapunov exponent obtained by synchronizing a numerical model to experimental data. The open squares were obtained by applying the same technique using numerically simulated time traces in place of experimental data. For comparison, the solid data points in Figure 4(a) show the largest Lyapunov exponent calculated by linearizing and discretizing the equations of motion, following the method of Farmer [14]. Unlike earlier methods that compute the Lyapunov exponent from numerical models, our method incorporates experimental measurements from a real system, which is essential if this technique is to be applied to the problem of prediction. This method is also unique in that it could provide a way of computing the local maximum Lyapunov exponents using two identical experimental systems even when the dynamical equations describing the system are not available.

Figure 4(b) shows a histogram of the prediction horizon, or inverse Lyapunov exponent, obtained for a feedback strength of $\beta = 4.00$ and input laser power $P = 675$ μW. For this feedback strength, the Lyapunov dimension was computed to be 15.6. The open circles show the distribution of prediction horizons determined from experimental data and the bars show the distribution obtained using numerically simulated time traces. The two distributions show a close correspondence, and the average Lyapunov exponents agree well with an independent calculation based on the linearized equations of motion. Whereas most existing numerical methods calculate only the global Lyapunov exponent of the system, the synchronization technique described here yields a statistical distribution of the finite-time Lyapunov exponent [11, 13].

Conventional methods of prediction, based on time-series analysis, prove difficult or impossible to implement for high-dimensional chaotic systems with time-delayed feedback. Such systems have been proposed and recently demonstrated for applications such as chaotic communication in fiber optic networks. We show that synchronization between two such systems (one of which may be a numerical model) is a powerful way to achieve data assimilation and drive the two systems to initially close starting points, from which future predictions can be made. It is remarkable that the prediction of high-dimensional dynamics can be made from observations of a scalar time series**.** By measuring the divergence of two initially synchronized systems, we can estimate the finite-time Lyapunov exponent – a key parameter that quantifies the local predictability of the system. The results show excellent agreement with independent numerical calculations and yield additional information about the distribution of finite-time Lyapunov exponents across the attractor. Prediction horizons are determined from experimental observations and simulations, establishing the applicability of this method to a wide variety of systems.

We acknowledge advice from Eric Forgoston, John Rodgers, Karl Schmitt and Ira Schwartz. Many of the ideas presented here arose at the 2008 Winter School for Hands-on Research on Complex Systems at the Institute for Plasma Research in Gandhinagar, India. This work was supported by DOD MURI grant (ONR N000140710734) and the US-Israel Binational Science Foundation.

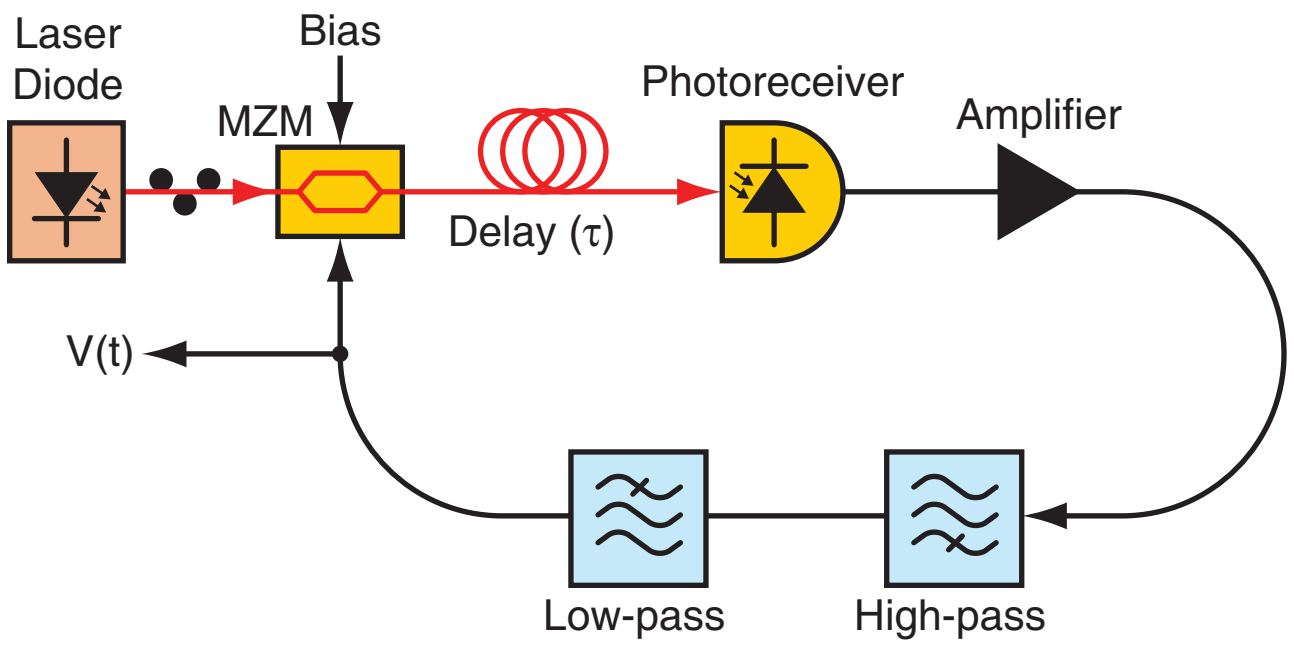

FIG. 1 (color online). The optoelectronic feedback loop studied in this work consists of a single frequency semiconductor laser, external Mach-Zehnder electrooptic modulator (MZM), photoreceiver, electronic amplifiers and filters that characterize the feedback system. The loop delay was 22.5 ns, and the electrical bandwidth of the loop is limited by the low-pass filter cut-off frequency of 100 MHz and the high-pass cut-on frequency of 1 MHz.

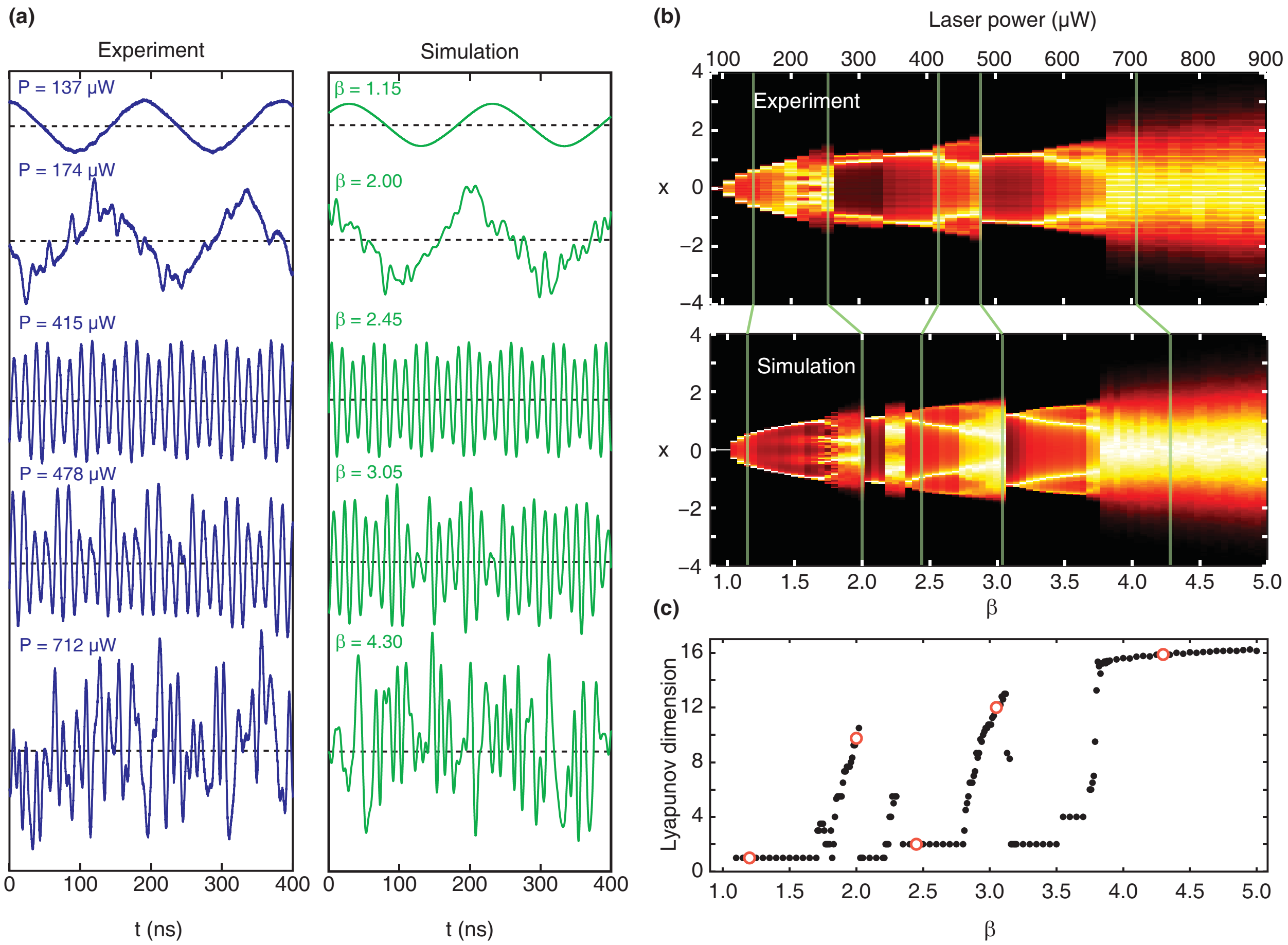

FIG. 2 (color online). Comparison of experimental observations and numerical simulations for varying feedback strength. In the experiments, the feedback strength is controlled by adjusting the laser power $P$, while in the simulations the normalized feedback strength $\beta$ is varied. Non-ideal behavior of the components in the system makes it difficult to establish a simple proportionality between $\beta$ and $P$ that holds for the entire range of dynamical behaviors. Nonetheless, we can see a close correspondence between the time traces and bifurcation sequences. (a) Measured and simulated time traces for five different values of the feedback strength. (b) Experimentally measured and numerically simulated bifurcation diagrams. (c) Lyapunov dimension as a function of feedback strength $\beta$, calculated using the Kaplan-Yorke conjecture [13] after computing the Lyapunov spectrum from a linearized numerical model [14]. In the fully chaotic range, on the right of the bifurcation diagram, the system exhibits approximately fifteen-dimensional dynamics. The vertical lines in (b) and open circles in (c) indicate the values of feedback corresponding to the five cases shown in (a).

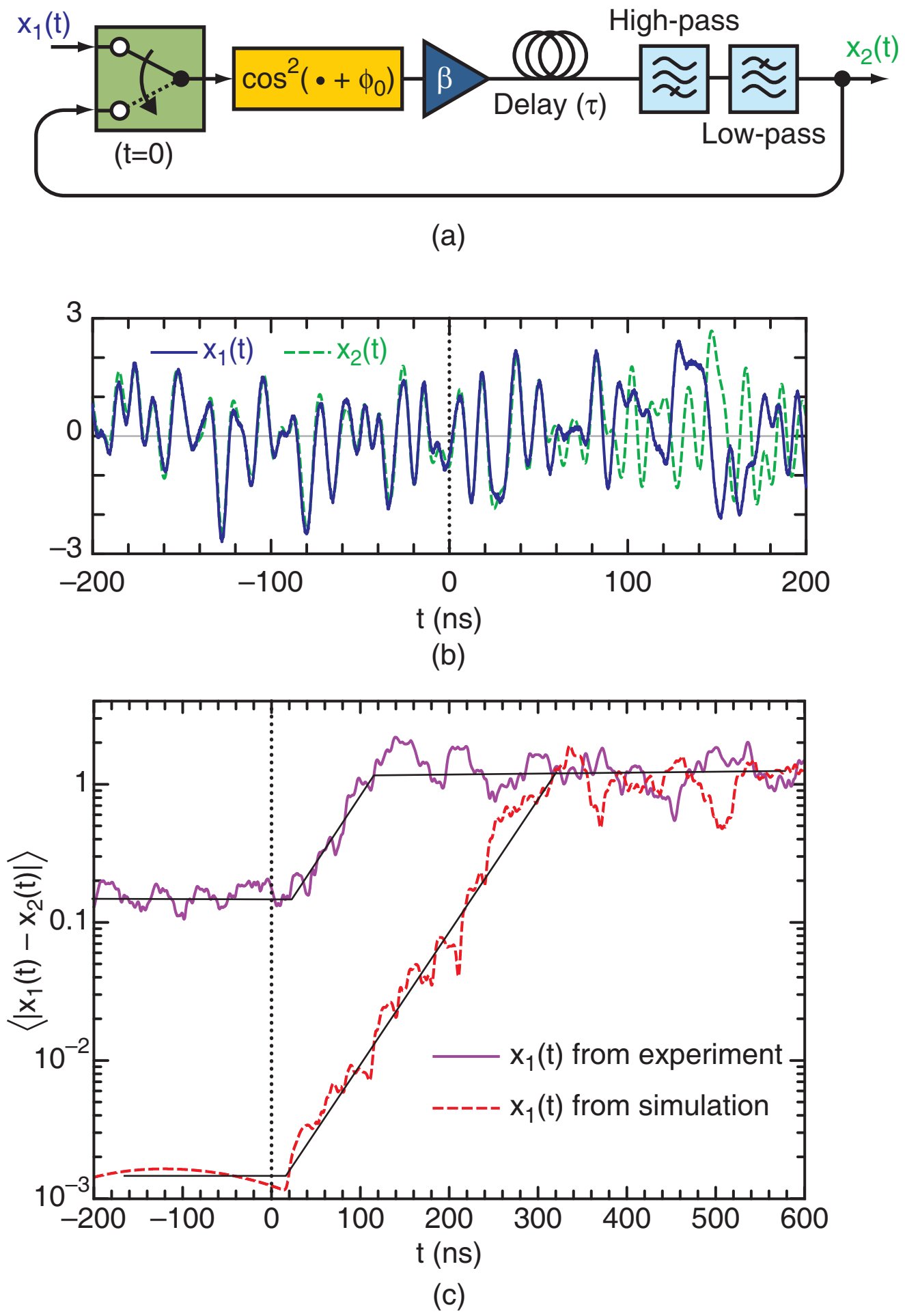

FIG. 3 (color online). Scheme for data assimilation and prediction. (a) Diagram illustrating the method by which a numerical simulation is synchronized to experimental observations. After the two signals $x_1(t)$ and $x_2(t)$ achieve synchrony, at $t = 0$ the switch is closed, which allows the numerical simulation to evolve independently. (b) Comparison between experimental and simulated time traces before and after the switch is closed, showing the divergence of the two waveforms. (c) Semilogarithmic plot of the absolute difference $|x_1 - x_2|$ showing the exponential divergence between the two waveforms after the switch is closed. The absolute difference is averaged using a 25 ns moving window for reliable estimation of the slope. The dashed curve was obtained by substituting a numerically simulated time trace for the experimental data, which yields much closer initial synchrony.

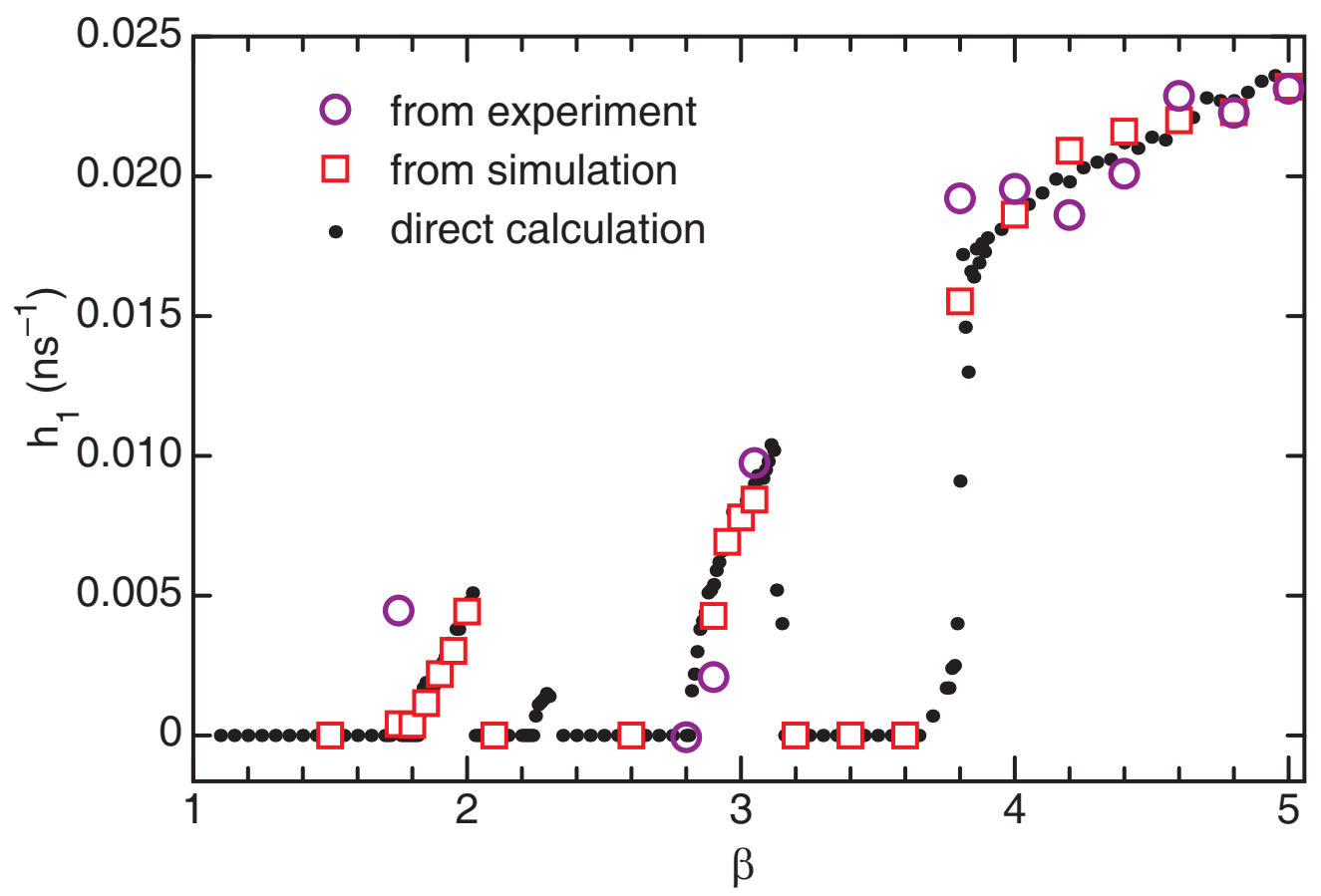

**(a)**

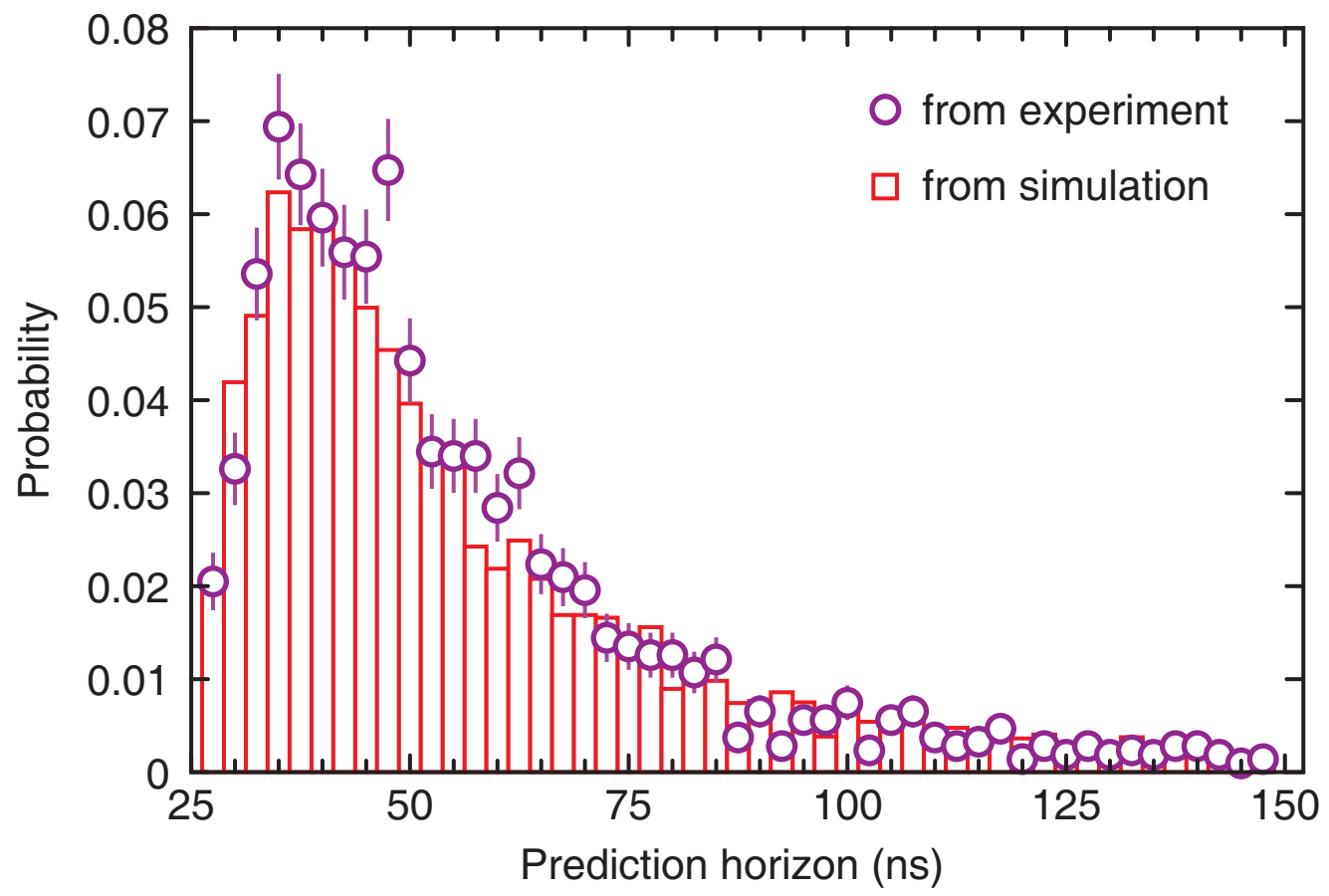

**(b)**

FIG. 4 (color online). Global maximal Lyapunov exponents and distribution of prediction times. (a) Average Lyapunov exponent $h_1$ as function of feedback strength $\beta$. The open circles show the average Lyapunov exponent obtained by synchronizing a numerical model to experimental data. The open squares were obtained by applying the same technique using simulated time traces in place of experimental data. The solid data points were calculated by linearizing and discretizing the equations of motion [14]. (b) Comparison of probability distribution of prediction horizon times when using experimental and simulated time traces at optical power $P$ = 675 μW and feedback strength $\beta$ = 4.00. The error bounds for the experimental distribution were calculated as the square root of the number of counts in each bin.